\documentclass[sigconf]{acmart}
\AtBeginDocument{%
  }

\setcopyright{acmlicensed}
\copyrightyear{2025}
\acmYear{2025}
\acmConference[GenAIECommerce 2025 @RecSys 25]{Workshop on Generative AI for E-Commerce}{September 22,
  2025}{Prague, Czech Republic}

\begin{document}

\title[Entropy driven dialogue policy]{Modeling shopper interest broadness with entropy-driven dialogue policy in the context of arbitrarily large product catalogs}

\author{Firas Jarboui}
\affiliation{
  \institution{Gorgias}
  \city{Paris}
  \country{France}
}

\author{Issa Memari}
\affiliation{
  \institution{Gorgias}
  \city{Paris}
  \country{France}
}


\begin{abstract}
  Conversational recommender systems promise rich interactions for e-commerce, but balancing exploration (clarifying user needs) and exploitation (making recommendations) remains challenging, especially when deploying large language models (LLMs) with vast product catalogs. We address this challenge by modeling the breadth of user interest via the entropy of retrieval score distributions. Our method uses a neural retriever to fetch relevant items for a user query and computes the entropy of the re-ranked scores to dynamically route the dialogue policy: low-entropy (specific) queries trigger direct recommendations, whereas high-entropy (ambiguous) queries prompt exploratory questions. This simple yet effective strategy allows an LLM-driven agent to remain aware of an arbitrarily large catalog in real-time without bloating its context window.
\end{abstract}



\received{10 July 2025}
\received[revised]{-}
\received[accepted]{30 August 2025}

\maketitle

\section{Introduction}
Recommender systems are increasingly turning to conversational agents to create interactive, personalized shopping experiences. Instead of passively presenting a list of items, a conversational recommender can ask questions, understand user preferences in real-time, and explain or refine recommendations through dialogue\cite{ren2021learning}. The recent breakthrough of large language models (LLMs) has accelerated this trend, as LLMs can fluidly generate natural language and handle complex user utterances\cite{feng2023large}.

By leveraging LLMs, conversational recommenders (CRS) have the potential to combine the strengths of traditional recommendation algorithms (database knowledge, collaborative filtering signals) with the flexible reasoning and language capabilities of AI assistants\cite{liu-etal-2023-conversational}. This synergy, however, comes with new challenges. Chief among them is how to manage the dialogue policy – i.e., deciding when the system should ask clarifying questions versus when it should present actual recommendations – in a way that takes advantage of the LLM’s capabilities without overwhelming it with information.

In this work, we propose an entropy-based routing strategy to address this challenge. The key idea is to quantitatively estimate the broadness or ambiguity of the user’s request by looking at the distribution of retrieval scores for candidate items. Intuitively, if the retrieval results are tightly clustered around a few high-scoring items, the conversational agent should exploit this and make a recommendation. Conversely, if the scores are spread out, it signals ambiguity – the agent should explore by asking a clarifying question or preference query.

We start by setting up the orchestration framework of the developed conversational AI before focusing on the its behavior as a shopping assistant. We then detail the product search pipeline, we discuss the limitations of naive retrieval augmented generation (RAG) and how that led us to develop the entropy-driven dialogue policy. We conclude the paper with empirical evidence supporting this work and the research directions we want to take in the future. Through this work, we aim to demonstrate that modeling user interest broadness via entropy is an effective and novel way to enhance LLM-driven recommender systems, contributing to more intelligent and adaptive conversational recommenders in practice.

\section{Multi–skill E-commerce AI Agent}
\label{sec:agent}

\begin{figure}[h]
  \centering
  \includegraphics[width=\linewidth]{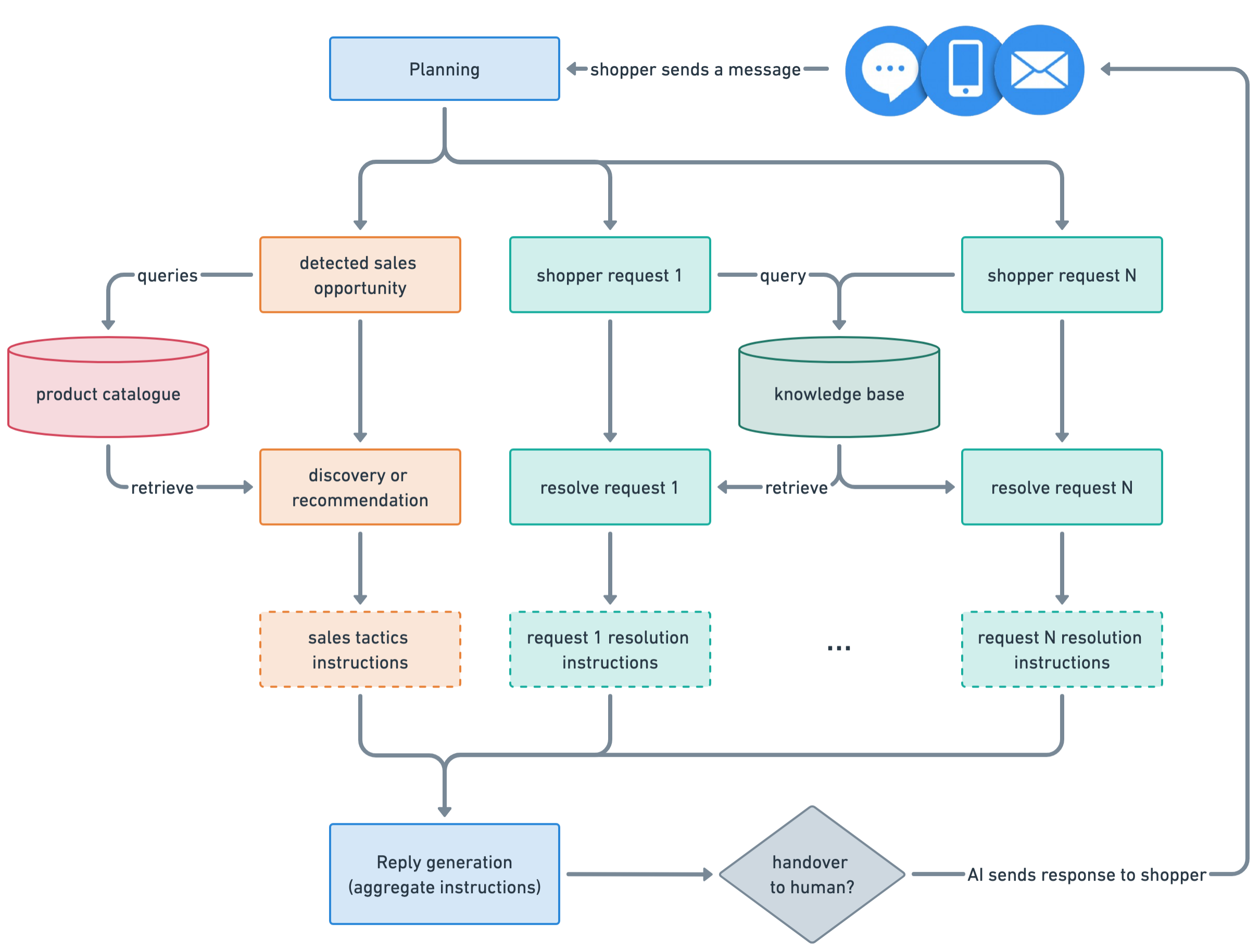}
  \caption{High-level data flow of the agent (see \S\ref{sec:agent}).}
  \label{fig:overview}
\end{figure}

We operate a production-grade AI agent that currently serves \textbf{${\sim}200\,\text{K}$ incoming shopper messages per day} across live chat widgets and e-mails for \textit{thousands} of e-commerce brands.
Merchants expect from the agent to \emph{both} remove purchase-blocking friction (e.g.\ “What’s your refund policy?”) \emph{and} actively increase basket conversion through persuasive, context-aware product recommendations\cite{zhang2018towards}.  
Treating “support” and “sales” in isolation would miss critical dependencies (clarifying a shipping deadline often opens a sales opportunity, and vice-versa) and would force costly hand-offs.  
We therefore build a \textbf{multi-skill agent} that unifies these tasks inside a single control loop (similar unified assistant strategies are described in Alibaba’s AliMe system for example \cite{li2017alime}).

Figure~\ref{fig:overview} schematizes the pipeline, which is organised into \textbf{three routing nodes}.  

\paragraph{\textbf{(1) Planning node:}}
Given the raw shopper message (and session metadata such as pages visited, cart contents, and prior orders), a large language model (LLM) decomposes the interaction into
\begin{itemize}
  \item \emph{support requests}—independent units of friction to be solved;
  \item a binary decision on whether a \emph{sales opportunity} exists.
\end{itemize}
The planner outputs a structured JSON plan that names each request and, if applicable, flags the turn for sales activation. This approach aligns with multi-intent detection in dialog systems, where a single utterance may contain multiple intents that need handling\cite{gangadharaiah-narayanaswamy-2019-joint}.

\paragraph{\textbf{(2) Resolution workflows:}}
Each element of the plan is processed by a dedicated \textbf{workflow}, i.e.\ a guarded chain of LLM calls with retrieval-augmented generation (RAG) over merchant data\cite{lewis2020retrieval}.

\begin{itemize}
  \item \textbf{Support-agent workflow} queries the merchant \emph{knowledge base} (policies, FAQs, etc.) and generates a concise resolution instructions. This ensures accurate answers to policy or support questions by retrieving relevant documents\cite{feng2020doc2dial}.
  \item \textbf{Shopping-assistant workflow} (focus of the remainder of this paper) selects a \emph{sales tactic}—\textit{discovery} or \textit{recommendation}—then produces tactic-specific \emph{instructions} that will later surface to the shopper. This design follows the conversational recommendation paradigm where the system can ask clarifying questions to discover user needs or present product recommendations when confident, as shown to improve recommendation relevance in interactive e-commerce settings\cite{zhang2018towards}.
\end{itemize}

\paragraph{\textbf{(3) Reply-generation node:}}
A final LLM aggregates all instruction digests, resolves overlaps, checks coherence, and produces a single reply.  
If any digest signals low confidence or excessive complexity, the node triggers a \emph{handover} to a human agent instead of auto-replying.

The remaining sections of this paper dive into the \emph{shopping-assistant workflow} and, in particular, the detailed mechanism that handle routing between discovery and recommendation tactics.

\section{AI Shopping Assistant}
\label{sec:assistant}

In every sales‐oriented turn, the agent must decide between two conversational \emph{tactics}:

\begin{enumerate}
  \item \textbf{Discovery} — ask clarifying questions or surface a small set of representative items to help the shopper refine their intent or explore the merchant’s offer.
  \item \textbf{Recommendation} — once the intent is precise, present a concrete product (or bundle) and explain how it satisfies the expressed need.
\end{enumerate}

Choosing the appropriate tactic at each turn is vital for both customer
experience and conversion \cite{ren2024information, jannach2021survey}. 
Implementing this mixed‐initiative behaviour raised two technical challenges:

\begin{description}
  \item[Product context retrieval] How can we fetch high‐quality candidate products without overflowing the LLM context window? We start by answering this in subsection \ref{sec:search} by detailing our two‐stage search pipeline and the distinction between our \emph{identification} and \emph{recommendation} retrieval operators.
  \item[Tactic selection.] How can the system infer turn by turn whether the dialogue is still exploratory or ready for a recommendation? The subsequent subsections review our baseline solution, its limitations, and how those limitations led us to the entropy‐based approach this paper introduces.
\end{description}

\subsection{Product Search System}
\label{sec:search}

Our shopping assistant relies on a two–stage neural search pipeline whose input is an \textbf{LLM-generated query} assembled from the live context—conversation turns, pages visited, cart contents, and past orders.  
Each query is executed on the merchant’s product catalogue in one of two modes:

\begin{itemize}
  \item \textbf{Identification}: retrieve \emph{matching} items to the expressed need.
  \item \textbf{Recommendation}: retrieve a \emph{cross-sell} or \emph{up-sell} items that complements the current interest.
\end{itemize}

\paragraph{\textbf{Retrieval architecture:}}
In a similar fashion to what prior research direction found to be relevant \cite{rossi2024relevance, li2021embedding, lin2024enhancing, freymuth2025hierarchical}, all the LLM generated queries and product descriptors are embedded with a \textbf{multilingual encoder} fine-tuned using a triplet loss (\textbf{\textsc{E5}} in our case\cite{wang2024multilingual}).  
We start by fetching candidates via an HNSW approximate-nearest-neighbour index\cite{malkov2018efficient}. Then we rerank the top-$k$ (here $k{=}50$) with a transformer-based\cite{vaswani2017attention} reranker model trained under \textit{binary cross-entropy} (BCE) on click-through labels, yielding a calibrated relevance score $s(q,p)$ used later for entropy computation (we used \textbf{RoBERTa} in our case\cite{liu2019roberta}).  

\paragraph{\textbf{Training data and negatives:}}
We exploit two complementary data sources to mitigate the exploration vs exploitation tradeoff of online recommender systems:

\begin{enumerate}
  \item \textbf{Out of distribution signals} collected organically without the AI agent intervention: 
  \begin{itemize}
      \item Identification: storefront search $\rightarrow$ landing-page pairs  
      \item Recommendation: cart co-occurrence tuples
  \end{itemize}
  \item \textbf{In distribution signals} success events collected from the live agent conversation: LLM-generated queries $\rightarrow$ identification or recommendation pipeline $\rightarrow$ click through events
\end{enumerate}

Negative examples are sampled uniformly from \textit{(i)} other merchants’ catalogues and \textit{(ii)} disjoint collections within the same merchant, ensuring hard negatives without semantic overlap.

\paragraph{\textbf{Cold-start robustness:}}
Because the encoder is trained on \textit{cross-merchant} data, it generalises to unseen shops; newly onboarded products are embedded on ingest and immediately indexed without additional fine-tuning—eliminating cold-start latency.

\subsection{Baseline: LLM-driven dialogue policy}
\label{sec:baseline}
At each sales focused round of interaction with the shopper, we define the shopping assistant dialogue policy as the decision to be made between: 
\begin{itemize}
    \item Asking clarifying questions or showing representative items to help refine the shopper interest (we will refer to this as the \textsc{Discovery} behavior)
    \item Being assertive and pushing the shopper toward purshasing a specific item (shopping assistant \textsc{Recommendation} mode) 
\end{itemize}

Our first generation shopping assistant relied on an LLM router to decide if we should go with either strategies. To achieve this, we combined \textbf{retrieval augmented generation (RAG)} with an LLM classifier. The workflow comprised three sequential phases:

\begin{enumerate}
  \item \textbf{Setup:}  Two parallel LLM calls are issued:  
        \begin{itemize}
          \item \emph{Interest classifier}—predicts the \textit{customer interest stage} as either \textsc{Discovery} (intent unclear) or \textsc{Interested} (intent precise);  
          \item \emph{Query generator}—produces one or more text queries conditioned on the full context (dialogue history, pages viewed, cart, etc.). These queries are targetting either the identification or the recommendation operators of our search engine.
        \end{itemize}
  \item \textbf{Candidate retrieval:}  Each query is executed by the two-stage search pipeline described in Section~\ref{sec:search}. 
  \item \textbf{Tactic prompting}  Depending on the predicted interest stage, the retrieved product snippets are injected into specialised prompts (either \textsc{Discovery} or \textsc{Recommendation}).  The prompt output is an \emph{instruction digest} forwarded to the reply-generation node.
\end{enumerate}

This baseline is very straight forward; nevertheless, it revealed important shortcomings.

\subsection{Observed Limitations}
\label{sec:limitation}

\paragraph{\textbf{Context-dependent granularity:}}
The LLM based interest stage classifier performed well when the shopper’s goal was \textit{orthogonal} to any specific product (“Can you ship before Friday?”). However, when the user explicitly mentioned a product or category, the classifier often misjudged whether the request was \emph{vague} or \emph{precise} \emph{relative to the merchant’s catalogue}. 

This problem is analogous to query quality performance in retrieval systems \cite{cronen2002predicting, hauff2008survey}. For instance, the query \textit{“nails”} is broad in a specialised nail-supply store (hundreds of SKUs) but highly specific in a general beauty store with only a few nail items.  Because the classifier does not have access to the full catalog breadth, it could not make this distinction.

A seemingly natural fix (providing the entire catalog as context for the LLM) is infeasible in practice as injecting thousands of products prohibitively increased hallucination risk. 
We could also add another intermediate prompt (between the \textbf{Candidate retrieval} and the \textbf{Tactic prompting} steps) but that would increase the first response time of AI agent, which is not desirable in practice as it increases the shopper drop off.  

\paragraph{\textbf{Shopping assistant discovery is shopper driven:}}
Now that we understand the limitations of the LLM-driven dialogue policy (deciding to go with a \textsc{Discovery} or a \textsc{Recommendation} behavior in the next round), we needed to quantify the missed opportunity addressable market among organic shopper interactions. For this, we evaluated the shopper intentions of their first message. 

\begin{figure}[h]
  \centering
  \includegraphics[width=\linewidth]{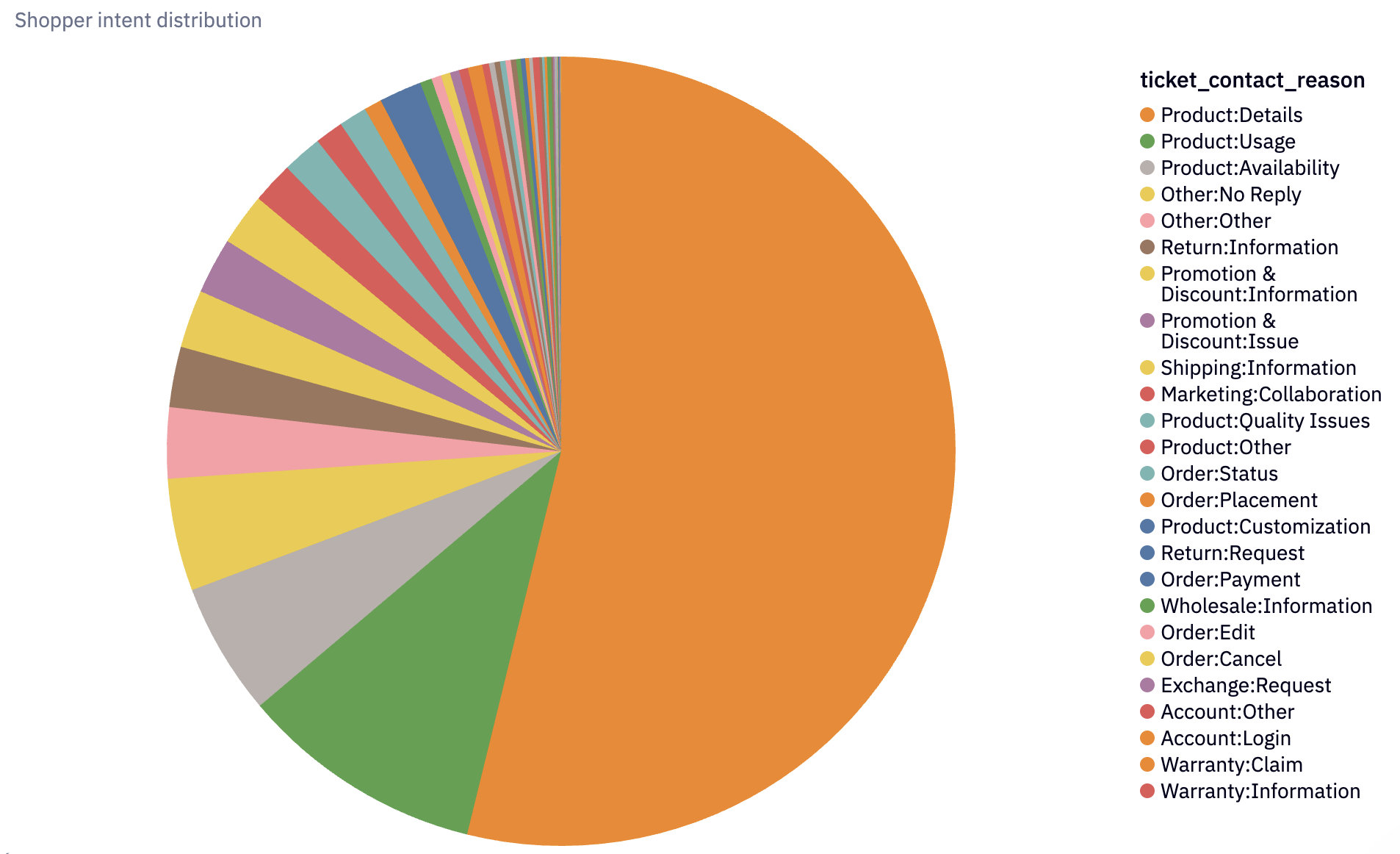}
  \caption{Distribution of shopper intents in e-commerce.}
  \label{fig:intents}
\end{figure}

We defined 52 intent categories and we refined a prompt to be aligned with a manually labeled balanced dataset of 50 examples of each category. We then used this prompt to infer the intent of 1M interactions sampled randomly over the last 6 months (January - June 2025). We report the distribution of the intents in Figure \ref{fig:intents}. The top 3 categories are \textsc{Product:Details}, \textsc{Product:Usage} and \textsc{Product:Availability}. They represent almost $\mathbf{70\%}$ of the full dataset ($\mathbf{54\%}$, $\mathbf{10\%}$, and $\mathbf{5.4\%}$ respectively).

These categories are basically indicating that almost always shoppers engage the AI with a clear product in mind. They are either asking about general information regarding the merchant catalog, specific details regarding the usage of a given product or its availability. 

Depending on the broadness of these queries (i.e. how precisely can we pin point a specific product based on the shopper context), the shopping assistant dialogue policy should behave very differently.
Intuitively, high quality queries and precise context with respect to the merchant catalog should lead to the \textsc{Recommendation} behavior, whereas vague queries should lead to the \textsc{Discovery} mode.

Looking at the scientific literature, we noticed that prior work investigated the correlation between query quality and the entropy of the similarity distribution\cite{zendel2023entropy}. This motivated our second-generation shopping assistant with \textbf{entropy-driven dialogue policy}, which leverages the \emph{distribution shape} of retrieved scores rather than shopper context to decide between \textsc{Discovery} and \textsc{Recommendation}.

\subsection{Modeling Query Breadth via Entropy}
\label{sec:entropy-metric}

Let the merchant catalogue contain \(N\) items \(\{\mathbf p_{1},\dots,\mathbf p_{N}\}\). For an LLM-generated query \(\mathbf q\) we obtain a calibrated re-ranker score \(s(\mathbf q,\mathbf p_i)\in[0,1]\) for each candidate product (Section~\ref{sec:search}). Over the top-\(k\) candidates we form a probability mass function
\[
  \mathbf{P}_i^q \;=\;
  \frac{s(\mathbf q,\mathbf p_i)}
       {\displaystyle\sum_{j=1}^{k}s(\mathbf q,\mathbf p_j)},
  \qquad i=1,\dots,k .
\]

Let \(\mathcal{D}_q\) be the distribution over the top-\(k\) retrieved
items for query \(\mathbf q\), where the probability of item \(i\) is $\mathbf{P}_i^q$

Shoppers arriving with very precise queries will produce a highly skewed distribution $\mathcal{D}_q$, with one or a few items taking most of the probability mass (very high similarity for a small set of products, and very low for the rest). In contrast, a shopper with a vague or broad-interest query yields a much flatter, uniform-like distribution of scores (many items have similarly low scores). Prior work has in fact quantified query ambiguity or specificity by measuring the entropy of the distribution of retrieval results or user clicks\cite{duan2012click}.

In our context, this motivates using the entropy of $\mathcal{D}_q$ as a proxy for the breadth of the customer’s intent. We define the \textbf{broadness score} as the normalized entropy of $\mathcal{D}_q$
\[
  B_k^N(\mathbf q)\;=\;
  \frac{-\sum_{i=1}^{k} P_i \log P_i}{\log k}
  \;\in\;[0,1].
\]

\begin{itemize}
  \item \(B_k^N(\mathbf q)=0\): \emph{minimum entropy}—all probability mass is concentrated on a single product, indicating \textbf{precise} intent.
  \item \(B_k^N(\mathbf q)=1\): \emph{maximum entropy}—scores are nearly uniform, signaling a \textbf{vague} query.
\end{itemize}

\paragraph{\textbf{Entropy estimator:}}
The broadness (or normalized entropy) of any given query $\mathbf{q}$, contextualized to the whole product catalog is $B_N^N(\mathbf q)$. In practice and for computational concerns, we only have access to the broadness contextualized to the top-\(k\) nearest retrieved items $B_k^N(\mathbf q)$. 

This means that the product catalogs items that will not be taken into account when computing the broadness are necessarily of lower similarity.
Subsequently, the top-\(k\) broadness $B_k^N(\mathbf q)$ will always over-estimate the full product catalog broadness score $B_N^N(\mathbf q)$ as we will not see the long tail of irrelevant items. 

\begin{figure}[h]
  \centering
  \includegraphics[width=\linewidth]{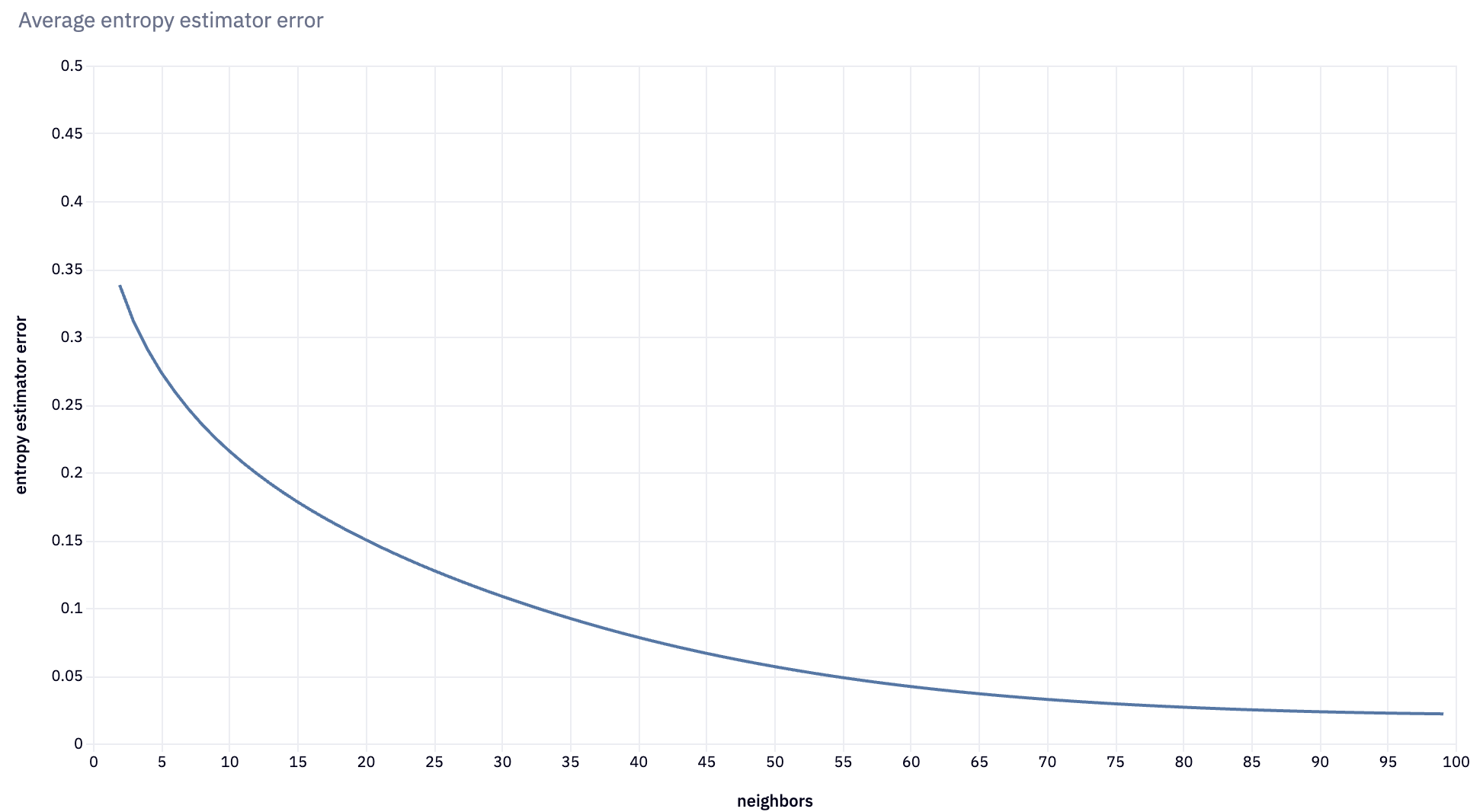}
  \caption{Convergence of the entropy estimator.}
  \label{fig:convergence}
\end{figure}

This begs the question: \textit{How far off is the top-\(k\) broadness score compared to the full product catalog contextualized one}. While we do not derive convergence bounds of this estimator in this paper, we investigated empirically the error over the organic search dataset (storefront search $\rightarrow$ landing-page pairs) averaged across all of Gorgias 15\textbf{K} supported stores. We report this in Figure \ref{fig:convergence}.

What we noticed is that average errors converges to 0 quite rapidly, and that with 50 neighbors we already have a reliable signal for our practical needs. 

\subsection{Entropy-Driven Dialogue Policy}
\label{sec:entropy-policy}

\begin{figure}[h]
  \centering
  \includegraphics[width=\linewidth]{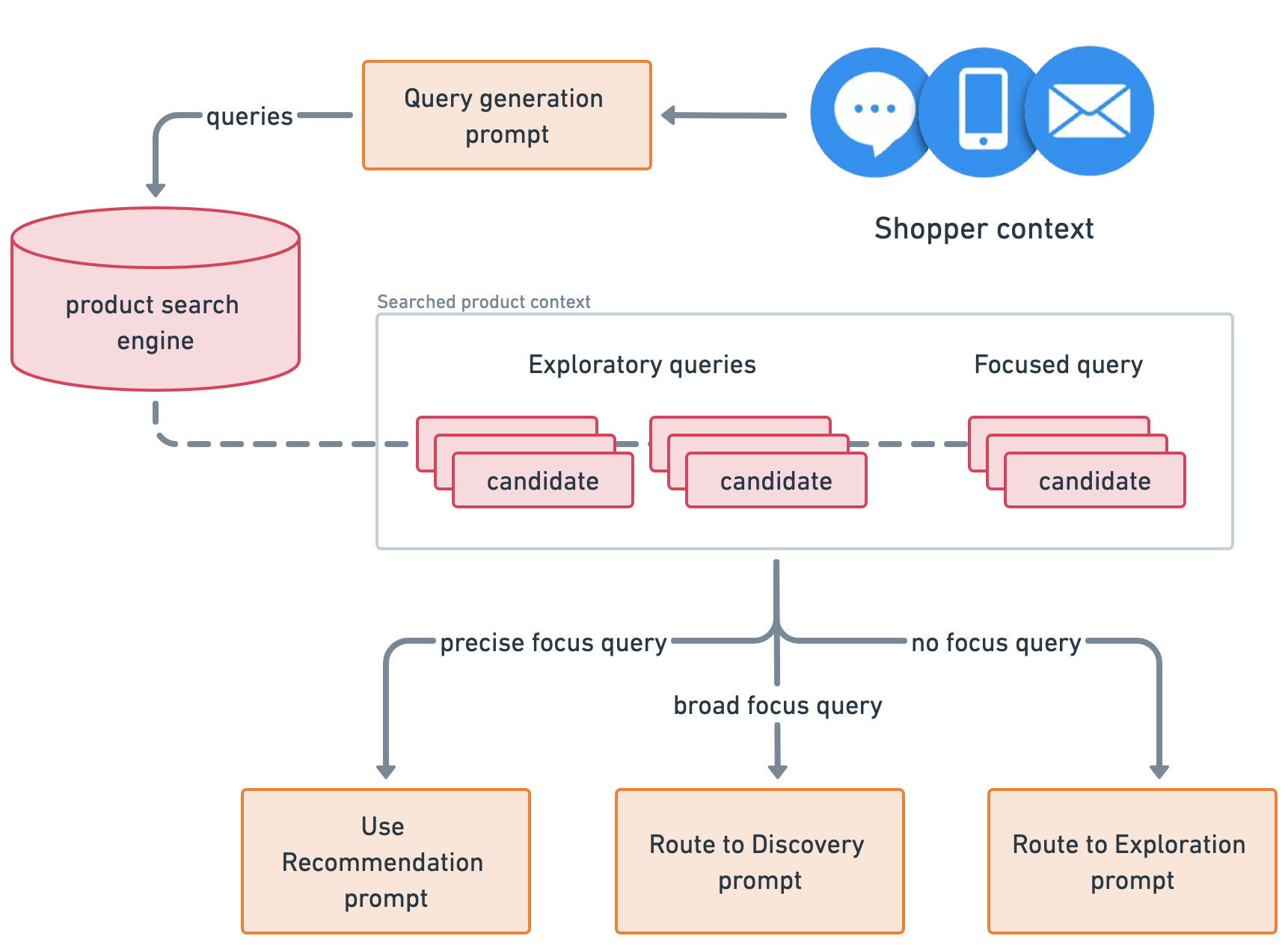}
  \caption{Entropy controlled sales tactic (see \S\ref{sec:entropy-policy}).}
  \label{fig:broadness}
\end{figure}

Using the broadness score, we design a simple entropy-sensitive dialogue policy for the shopping assistant. At each dialogue turn, a query generator LLM defines two types of search queries based on the user’s utterance and context:

\begin{description}
  \item[Exploratory] one or more catalog-level identification or recommendation searches that reflect the user’s general context (pages viewed, cart contents, dialogue history).
  \item[Focused] a single identification query distilled from the most recent utterance. This is our best guess at the shopper’s immediate target.
\end{description}

We execute all candidate queries through the product search engine, and then compute the broadness score for the focused query \(B_{\text{focused}}\). The dialogue policy then branches based on the presence of a confident focused query and its entropy value:

\begin{itemize}
  \item \textbf{No focus query} If the LLM cannot produce a plausible focused search, we route the conversation to an Exploration phase. The system aims to pique the shopper’s interest by showcasing a diverse set of contextually relevant products or asking a broad question. The assistant might show popular categories or related items to help the user articulate their needs.
  \item \textbf{Focus query present}:  
        \begin{itemize}
          \item \(B_{\text{focus}} < \tau_{\text{merchant}}\): 
                This indicates low entropy (highly concentrated relevance on a few items) – the user’s intent appears precise. The assistant will confidently suggest the top result or a very narrow set of products that are most likely relevant to the shopper’s need. This corresponds to a high-confidence scenario where exploitation is favored over exploration. 
          \item \(B_{\text{focus}} \ge \tau_{\text{merchant}}\)
                This indicates high entropy (a broad or ambiguous query) – the user’s intent is still vague or underspecified. The assistant will ask one or two clarifying questions or request additional details from the shopper. The goal is to reduce the entropy in the next turn by encouraging the user to narrow down their requirements. 
        \end{itemize}
\end{itemize}

An important advantage of this entropy-driven policy is that it remains continuously catalog-aware without expensive context injection. The decision logic depends only on the distribution of retrieval scores, which means any newly indexed product or change in the catalog can immediately influence the outcome in real-time.

\paragraph{\textbf{Threshold selection:}}
The cutoff level at which we switch between strategies controls a trade-off between how many clarifying questions the shopping assistant would ask during discovery and how likely are we to identify the actual product of interest for the shopper when we move into the recommendation mode.

The threshold \(\tau_{\text{merchant}}\) is bucketed into three aggressiveness presets (\emph{educational}, \emph{balanced}, \emph{pushy}), allowing merchants to trade off depth of discovery against conversion pressure. These presets are experimental and can be refined, but they illustrate how entropy-based routing can flexibly support different sales tactics.

\begin{figure}[h]
  \centering
  \includegraphics[width=\linewidth]{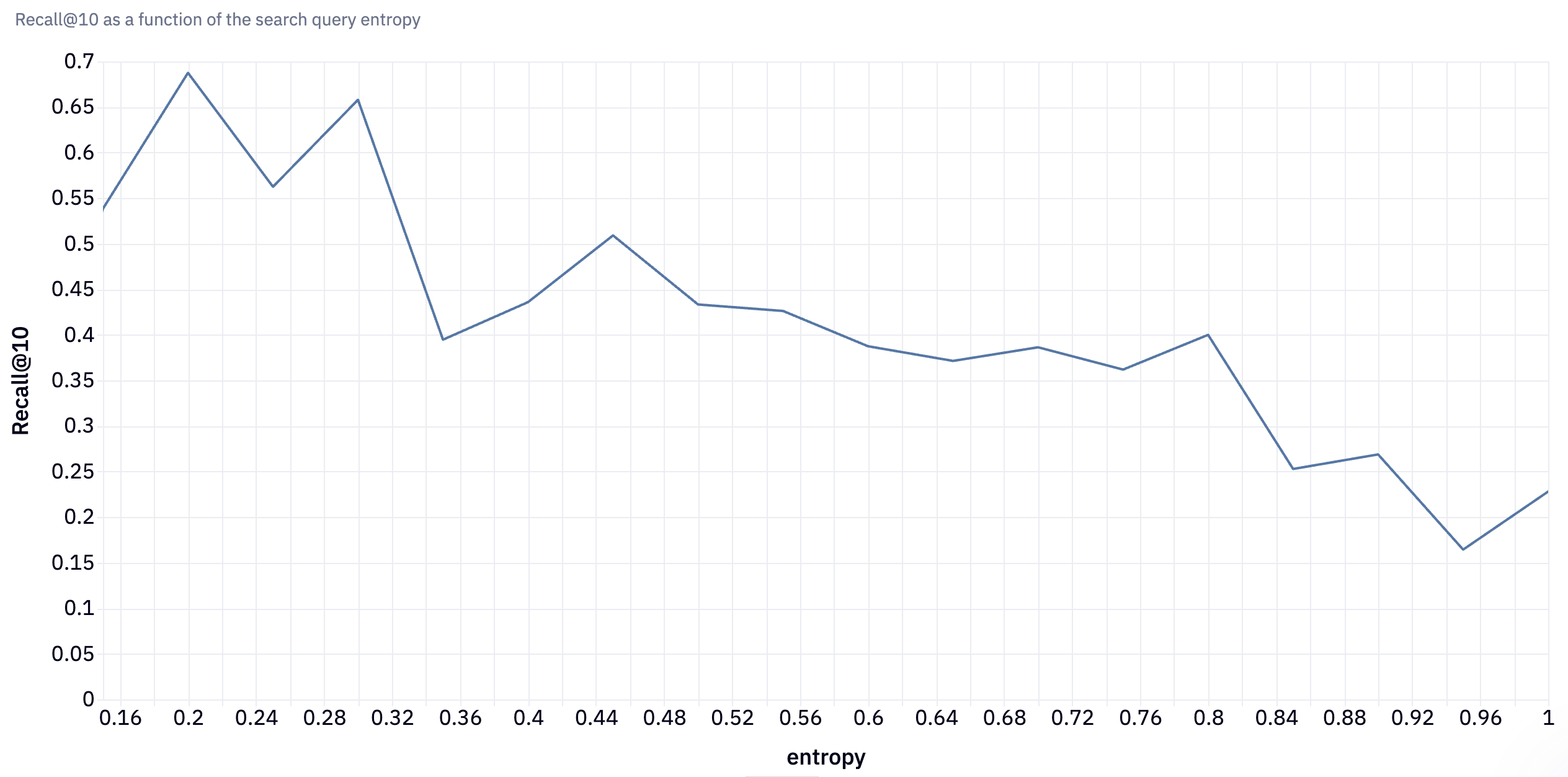}
  \caption{Recall at 10 plateau for different values of entropy}
  \label{fig:recall_entropy}
\end{figure}

In order to define a the levels of each bucket, we used the organically collected user search queries from the shoppers session data which were mapped into the next landing page after the search. For each query, we measure the recall at 10 our retrieval system (binary, 0 or 1) and we measure the entropy of the search. 

We then measured the average normalized entropy of search queries of same entropy buckets, and we report the average recall at 10 as a function of the entropy bins if Figure \ref{fig:recall_entropy}.

Looking at the data we noticed that recall is showing plateaus of similar recall at 10 before dropping around the values (0.3 and 0.8):
\begin{itemize}
    \item \textbf{lower than 0.3:} we have a probability of around 0.6 of retrieving the correct product item among the top 10 retrieved ones
    \item \textbf{between o.3 and 0.8:} the probability is stable around 0.4
    \item \textbf{beyond 0.8:} the success here drops to 0.2 for identifying the next landing product page for queries with such broadness score
\end{itemize}

We used these levels as an educated guess for the threshold \(\tau_{\text{merchant}}\) of the three aggressiveness presents mentionned above.

\section{Impact on engagement and future research}

\begin{figure}[h]
  \centering
  \includegraphics[width=\linewidth]{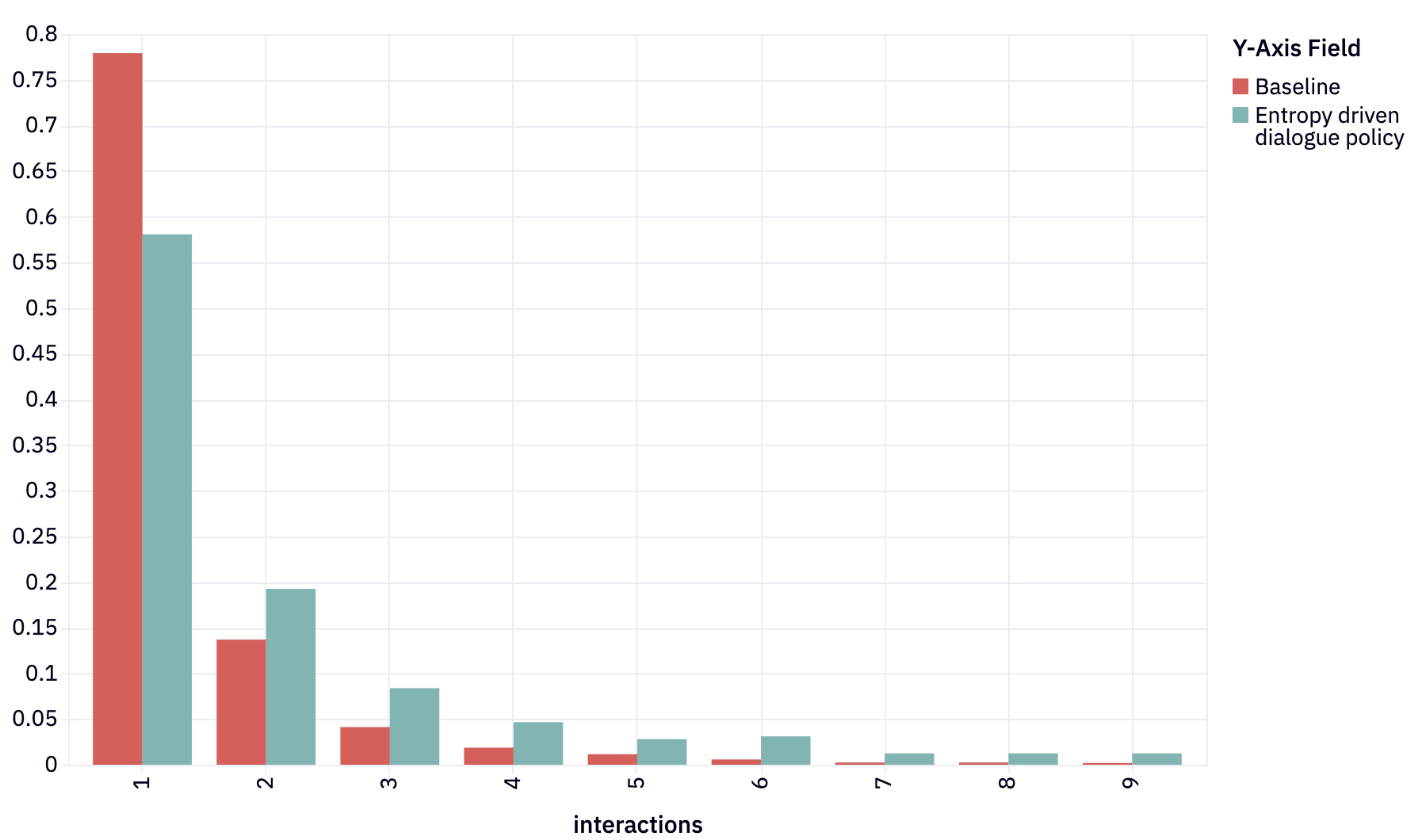}
  \caption{The shopping assistant powered with the entropy driven dialogue policy had longer conversations with shoppers than our baseline}
  \label{fig:length}
\end{figure}

The goal of this direction of work was to increase the shopper engagement with the shopping assistant as a success metric. We define engagement as the average number of rounds of interactions a shopper has with the AI during the same session. 

We ran an AB test on real customers to evaluate how different dialogue policies affect the conversation length, the distribution of conversation length is reported in Figure \ref{fig:length}.

We can clearly see that the entropy driven policy led to more engaging conversations with the shopper as the shopping assistant was not pushing items for sale but trying to collect more context from the user to give them a better recommendation.

We measured during the AB test whether this affected the conversion rate of conversations, but while we noticed a slight increase, it remained statistically insignificant to be reported in these results. We believe that there are way more factors involved in affecting the conversion rate for the dialogue policy to have a direct effect on it. 
This work merely covers one aspect of the complex dynamic involved with the conversational recommender system that interacts with live shoppers. 

A future research direction that we are actively looking into is how to make the shopping assistant proactive while minimizing the risk of distrcting the end user (instead of waiting for the customer to engage the AI, devise a policy that would review the shopper browsing data and context to decide if the shopping assistant should offer help organically).

\begin{acks}
We thank \textbf{Gorgias} for supporting this research and enabling its deployment at scale. We also wish to acknowledge the engineering and analytics contributors behind the implementation of the shopping assistant, whose work streamlined the measurement of conversational strategies in production and made it possible to rigorously evaluate entropy-based dialogue policies in live shopper interactions.
\end{acks}

\bibliographystyle{ACM-Reference-Format}
\bibliography{Content/biblio}

\end{document}